\documentclass[prb,twocolumn,showpacs,floatfix]{revtex4}
\usepackage{graphics}
\usepackage{graphicx}
\usepackage{amssymb}
\begin{document}
\newcommand{\RR}{\mathrm{\mathbf{R}}}
\newcommand{\rr}{\mathrm{\mathbf{r}}}
\newcommand{\defin}{\stackrel{def}{=}}

\title{Electron gas at the interface between two antiferromagnetic insulating manganites}

\author{M.J. Calder\'on}
\affiliation{Instituto de Ciencia de Materiales de Madrid (CSIC), Cantoblanco, 28049 Madrid (Spain)}
\author{J. Salafranca}
\affiliation{Instituto de Ciencia de Materiales de Madrid (CSIC), Cantoblanco, 28049 Madrid (Spain)}
\author{L. Brey}
\affiliation{Instituto de Ciencia de Materiales de Madrid (CSIC), Cantoblanco, 28049 Madrid (Spain)}
\date{\today}

\begin{abstract}
We study theoretically the magnetic and electric properties of the interface between two antiferromagnetic and insulating manganites: La$_{0.5}$Ca$_{0.5}$MnO$_3$, a strong correlated insulator, and CaMnO$_3$, a band-insulator. We find that a ferromagnetic and metallic electron gas is formed at the interface between the two layers.  We confirm the metallic character of the interface by calculating the in-plane conductance. The possibility of increasing the electron gas density by selective doping is also discussed.
\end{abstract}
\maketitle
\section{Introduction}

\begin{figure}
\resizebox{85mm}{!}{\includegraphics{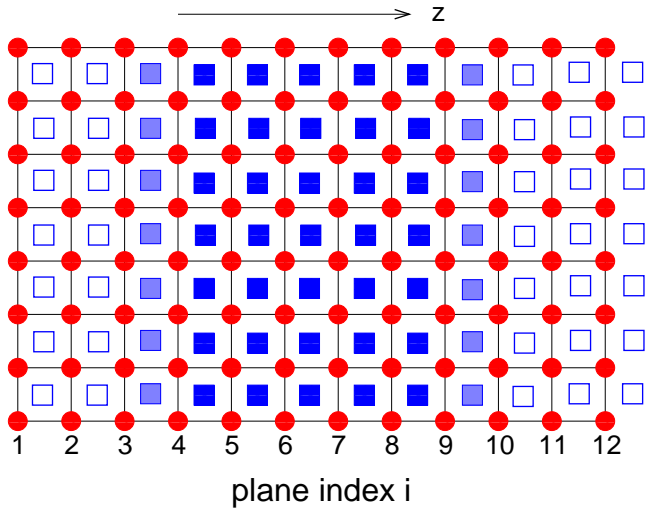}}
\caption{\label{fig:lattice} (Color online) 2-dimensional (xz plane) projection of the cubic lattice considered. The circles represent the Mn sites while the squares are the A-sites (Ca or La) shifted by (1,1,1)$\frac{a}{2}$ with respect to the Mn. $a$ is the lattice parameter. Full squares represent the La$_{0.5}^{3+}$Ca$^{2+}_{0.5}$O$^{2-}$ plane, with a charge density of $+0.5$ per A atom. Empty squares correspond to the Ca$^{2+}$O$^{2-}$ plane, which is neutral. Between, the gray (light blue) squares correspond to the average $+0.25$. We consider periodic boundary conditions in all three directions.
}
\end{figure}

In recent years, atomic control has been gained in the growth of multilayers of thin oxide films with different electronic and magnetic properties.~\cite{ohtomo02} Interestingly, 'electronic reconstruction' (or redistribution of charges) at the interfaces of strongly correlated systems~\cite{okamoto-nat04} has led to the appearance of phases showing a very different behavior from that in the bulk of the constituent materials.~\cite{ohtomo02,okamoto-nat04,ohtomo04,okamoto05,pena05,lee06,brinkman07,reyren07,gonzalez08}. 
For example, superlattices of thin layers of LaTiO$_3$ and SrTiO$_3$ show metallic conductivity, although both materials are insulators in bulk.~\cite{ohtomo02} SrTiO$_3$ is a band insulator while LaTiO$_3$ is a Mott insulator and has one extra electron on the Ti. This extra charge spreads across the interface producing the metallic behavior.~\cite{okamoto-prb04} 
Another interesting example is the formation of a high mobility electron gas at the interface between the two band insulators LaAlO$_3$ and SrTiO$_3$.~\cite{ohtomo04} This electron gas has been shown to have magnetic properties~\cite{brinkman07} and  
to behave as a two dimensional superconductor at low temperatures.~\cite{reyren07} The electric field effect, which is the basis for semiconductor transistors, can also be used to tune the charge density at interfaces between strongly correlated oxides, changing their properties.~\cite{ahn-nat03}

The redistribution of charge in Mott-band insulator heterostructures and the formation of a metallic interface have been studied theoretically using different techniques including Hartree-Fock theory to account for the on-site correlations,~\cite{okamoto-nat04,okamoto-prb04} dynamical mean field theory,~\cite{okamoto-PRB04R} and Lanczos diagonalization of a quasi-one-dimensional lattice.~\cite{kancharla06} The common observation is the formation of a metallic interface region up to three-unit cells wide. Interfaces between two different Mott insulators have also been theoretically studied:~\cite{lee06} a two-dimensional metal is formed with a density of states that can be controlled by a remote (a few unit-cells away) doping layer.

Perovskite manganites of formula A$^{3+}_n$A'$^{2+}_{1-n}$MnO$_3$ (with $n$ the density of electrons) are strongly correlated magnetic materials that present a great variety of phases depending very strongly on the close competition between different interactions: insulating behavior is favored by electron-lattice interaction, Coulomb interaction, and antiferromagnetic superexchange, while ferromagnetism and metallicity are favored by the double-exchange mechanism.~\cite{dagotto-book} By simply changing the doping of a manganite you can go from ferromagnetic and metallic phases to antiferromagnetic and insulating. Phase separation is also commonly observed.~\cite{dagotto-book,mathur03,israel07}  All-manganite heterostructures are therefore a very interesting ground for analyzing the interplay between different phases, while minimizing the creation of defects at the interfaces  thanks to the similarity in the lattice structure and chemical composition of the layers.~\cite{li02,lin06,brey-PRB07,niebieskikwiat07,salafranca08}

Motivated by the 'electronic reconstruction' effects observed in different oxide heterostructures and by the range of different phases that manganites show depending on their doping level, we have studied an all-manganite heterostructure formed by two different manganite insulators: La$_{0.5}$Ca$_{0.5}$MnO$_3$ (LCMO) and CaMnO$_3$ (CMO). LCMO is a CE-type antiferromagnet (consisting of ferromagnetic zig-zag chains in the x-y plane coupled antiferromagnetically to each other) and insulator in which the antiferromagnetic (AF) superexchange and the Jahn-Teller interaction open a gap in the band structure. On the other hand, CMO is a parent compound with no $e_g$ electrons on the Mn sites and, therefore, a band insulator. CMO presents AF ($\pi,\pi,\pi$) ordering (G-type) due to the superexchange between the $t_{2g}$ electrons.

By looking at the band structure and the in-plane conductance of these heterostructures, we find that a ferromagnetic (FM) and conducting interface can be formed for a range of reasonable parameters. We also find that by changing the electron concentration $n$ in LCMO around half-doping, $n=0.5\pm\delta$, the charge in the metallic interface and the in-plane conductance can be manipulated. 

This paper is organized as follows. In Sec.~\ref{sec:model} we describe the model we use to describe the manganite heterostructures. The phase diagram, the characterization of the heterostructure ground state configuration, and the effect of doping around $n=0.5$ are shown in Sec.~\ref{sec:results}. Finally, we conclude in Sec.~\ref{sec:conclusion}.

\section{Model}
\label{sec:model}

We study the heterostructure depicted in Fig.~\ref{fig:lattice} consisting of alternating CMO and LCMO layers a few unit cells wide. The background charge (namely, the charge on the AO planes) is $0$ in the CMO planes, $+0.5$ in the LCMO planes and, to achieve a symmetric distribution of the charge, we use the average $+0.25$ at the interfaces. To study the phase diagram, different magnetic configurations on the CMO planes (either FM or G-type AF) are considered. On the other hand, the magnetic order in the LCMO planes is always considered to be of the CE-type: FM zig-zag chains in the x-y plane coupled antiferromagnetically with neighboring chains. As discussed below, FM planes in the LCMO layer are not favored energetically.

The hamiltonian~\cite{brey-PRL04,brey-PRB07,salafranca08} includes the kinetic energy, a nearest neighbor antiferromagnetic coupling, and a Hartree term that takes into account the long range Coulomb interaction between all the charges in the system

\begin{equation}
H =- \sum_{i,j,\gamma,\gamma'} f_{i,j} t^u_{\gamma,\gamma'} C_{i,\gamma}^{\dagger} C_{j,\gamma'} + \sum_{i,j}  J_{\rm AF}^{ij} {\mathbf S}_i  {\mathbf S}_j  + H_{\rm Hartree} \, ,
\label{eq:H}
\end{equation}

where $C_{i,\gamma}^{\dagger}$ creates an electron on the Mn i-site, in the $e_g$ orbital $\gamma$ ($\gamma=1,2$ with $1=|x^2-y^2 \rangle$ and $2=|3 z^2-r^2 \rangle$). The hopping amplitude $f_{i,j}$ depends on the Mn core spins orientation given by the angles $\theta$ and $\psi$ via   the double-exchange mechanism 
{\small
\begin{equation}
f_{i,j}=\cos(\theta_i/2) \cos(\theta_j/2)+ \exp[i(\psi_i-\psi_j)]\sin(\theta_i/2) \sin(\theta_j/2)  \, , 
\end{equation}
}
and on the orbitals involved $t^{\rm x(y)}_{1,1}=\pm \sqrt{3} \,t^{\rm x(y)}_{1,2}=\pm \sqrt{3}\, t^{\rm x(y)}_{2,1}=3\, t^{\rm x(y)}_{2,2}=3/4 \,t^{\rm z}_{2,2}=t$ where the superindices x,y, and z refer to the direction in the lattice. All the parameters are given in units of $t$ which is estimated to be $\sim 0.2 - 0.5$ eV. $J_{\rm AF}$ is, in general, an effective antiferromagnetic coupling between first neighbor Mn core spins which is different in the CMO and LCMO layers. For CMO, the antiferromagnetic coupling is a pure superexchange between the localized t$_{2g}$ spins, while for LCMO it would effectively include the interaction between the e$_g$ electrons and the lattice.~\cite{vandenbrink-PRL99}
$H_{\rm Hartree}$ takes the form
{\small
\begin{equation}
H_{\rm Hartree} ={{\frac{e^2}{\epsilon}}} \sum_{i \ne j} \left({\frac{1}{2}} {\frac {\langle n_i \rangle   \langle n_j \rangle}{|{\mathbf R}_i-{\mathbf R}_j |}} +{\frac{1}{2}} {\frac {Z_i Z_j}{|{\mathbf R}^A_i-{\mathbf R}^A_j |}}
-{\frac{Z_i   \langle n_j \rangle}{|{\mathbf R}^A_i-{\mathbf R}_j |}}\right)
\label{eq:hartree}
\end{equation}
}%
with ${\mathbf R}_i$ the position of the Mn ions, $\langle n_i \rangle=\sum_{\gamma} \langle  C_{i,\gamma}^{\dagger} C_{i,\gamma} \rangle$ the occupation number on the Mn i-site, $eZ_i$ the charge of the A-cation located at ${\mathbf R}_i^A$, and $\epsilon$ the dielectric constant of the material. The relative strength of the Coulomb interaction is given by the parameter $\alpha=e^2/a \epsilon t$~[\onlinecite{lin06}].

We are assuming that the e$_g$ Mn levels in the two different layers are aligned, although a band offset due to a mismatch in the work functions of the two materials could occur.~\cite{lee06} We expect our approximation to be reasonable as the two layers are chemically similar. 

We find the ground state configuration (at temperature $T=0$) of the heterostructure by solving the hamiltonian in Eq.~\ref{eq:H} self-consistently in a $4 \times 4 \times 12$ system ($6$ unit cells of CMO and other $6$ of LCMO, as depicted in Fig.~\ref{fig:lattice}) with periodic boundary conditions in the three directions. Bulk behavior is always recovered at the center of the layers and the two CMO/LCMO interfaces (at planes $i=3-4$ and planes $i=9-10$) are independent of each other.

\section{Results}
\label{sec:results}

\subsection{Phase diagram} 
\label{subsec:pd}

\begin{figure}
\resizebox{70mm}{!}{\includegraphics{Fig2.eps}}
\caption{\label{fig:pd} (Color online) Phase diagram for $n=0.5$ in terms of the two parameters $\alpha$ and $J_{\rm AF}$. Results are very similar for different values of $n$. The label in each region stands for the number of FM planes in CMO at the interface with LCMO. The 3FM region corresponds to a FM-CE heterostructure. In a wide range of reasonable values of $J_{\rm AF}$ (see text), the configuration with one FM plane between G and CE ($1$FM) is stable. 
}
\resizebox{75mm}{!}{\includegraphics{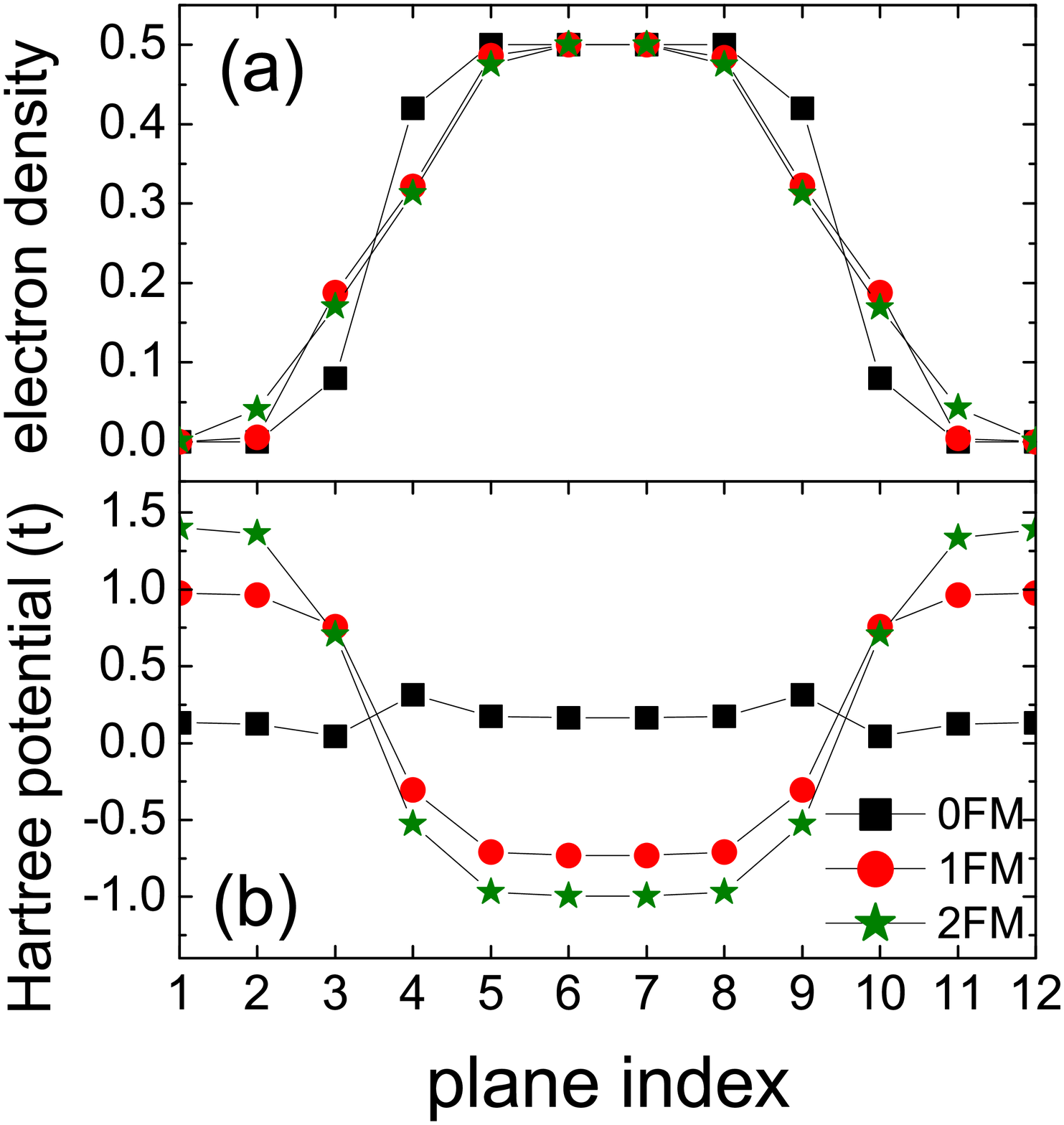}}
\caption{\label{fig:cpl-0FM-1FM} (Color online) (a) Electron density at each plane for $n=0.5$ and $\alpha=2$ for the configurations with no FM plane at the interface ($0$FM), with one FM plane at the interface ($1$FM), and with two FM planes ($2$FM). The curves for $3$FM are very similar to $2$FM. This illustrates how the electronic charge spreads across the interface. This spreading is limited by the Hartree potential shown in (b).
}
\end{figure}

In Fig.~\ref{fig:pd} we show the phase diagram of the heterostructure built by comparing the energies of four different configurations in the CMO layer. All the considered configurations have a symmetry plane between planes $i=6$ and $7$ in Fig.~\ref{fig:lattice}: (i) the configuration labeled $0$FM corresponds to G-type AF order on all CMO planes ($i=1$ to $3$, and $i=10$ to $12$); (ii) $1$FM stands for one FM plane at each interface ($i=3$ and $10$); (iii) $2$FM stands for two FM planes at each interface ($i=2$ and $3$ on one interface, and $i=10$ and $11$ on the other); and, finally, (iv) $3$FM corresponds to all FM CMO. We have checked that having FM planes in the CMO layer is more favorable energetically than having FM planes in the LCMO layer. This is mainly due to the fact that the Jahn-Teller coupling occurring in bulk LCMO, and absent in bulk CMO (with no e$_g$ electrons), implies a larger effective AF superexchange in the former,~\cite{vandenbrink-PRL99} making AF order more robust in LCMO than in CMO. 

As the LCMO layer has a fixed CE configuration in all planes, the phase diagram is independent of the $J_{\rm AF}^{\rm LCMO}$ coupling and it only depends on $J_{\rm AF}^{\rm CMO}$ which is a pure AF superexchange ($J_{\rm AF}^{CMO}=J_{\rm AF}\sim 1-10$ meV~[\onlinecite{dagotto-book}]). An estimation of $J_{\rm AF}$ in units of the hopping parameter $t=0.1-0.5$ eV would be $J_{\rm AF}\sim 0.02 - 0.1 \, t $. In Fig.~\ref{fig:pd} we plot the phase diagram as a function of $\alpha$ and $J_{\rm AF}$. A likely value for $\alpha$ is in the range $\sim 1-2$~[\onlinecite{lin06}]. 

In Fig.~\ref{fig:cpl-0FM-1FM} we show the electronic density per plane and the Hartree potential for the $0$FM, $1$FM, and $2$FM configurations. The results are shown for $\alpha=2$ (the curves are qualitatively similar for different values of $\alpha \sim 1-2$). Fig.~\ref{fig:cpl-0FM-1FM} illustrates the charge spreading across the interface and how it is limited by the Coulomb interaction between all the charges in the system, given by the average on each plane of the Hartree potential, which at site $i$ takes the form
\begin{equation}
V_{\rm Hartree}(i)= {\frac{e^2}{\epsilon}} \sum_{j\neq i} \left({\frac{\langle n_j \rangle}{|{\mathbf R}_i-{\mathbf R}_j |}}-{\frac{Z_j}{|{\mathbf R}_i-{\mathbf R}_j^A |}} \right) \,\, .
\end{equation}
The bulk values for the electronic density are recovered in the center of both CMO and LCMO layers. The curves for the $3$FM (not shown) are almost coincident to the ones corresponding to $2$FM.  

The phase diagram (Fig.~\ref{fig:pd}) shows that configurations with ferromagnetic planes are the ground state for the lowest values of the $J_{\rm AF}$. For reasonable values of this parameter, the ground state is the $1$FM configuration. From now on, we focus on this particular case.

\subsection{Analysis of the metallic interface: density of states, conductance, and 'shadow' order.}
\label{subsec:analysis}

\begin{figure}
\resizebox{65mm}{!}{\includegraphics{Fig4.eps}}
\caption{\label{fig:dosperfect} (Color online) Density of states (DOS) for a perfect CE antiferromagnet (a) and a 2-dimensional FM (b).
}
\vspace{1cm}
\resizebox{85mm}{!}{\includegraphics{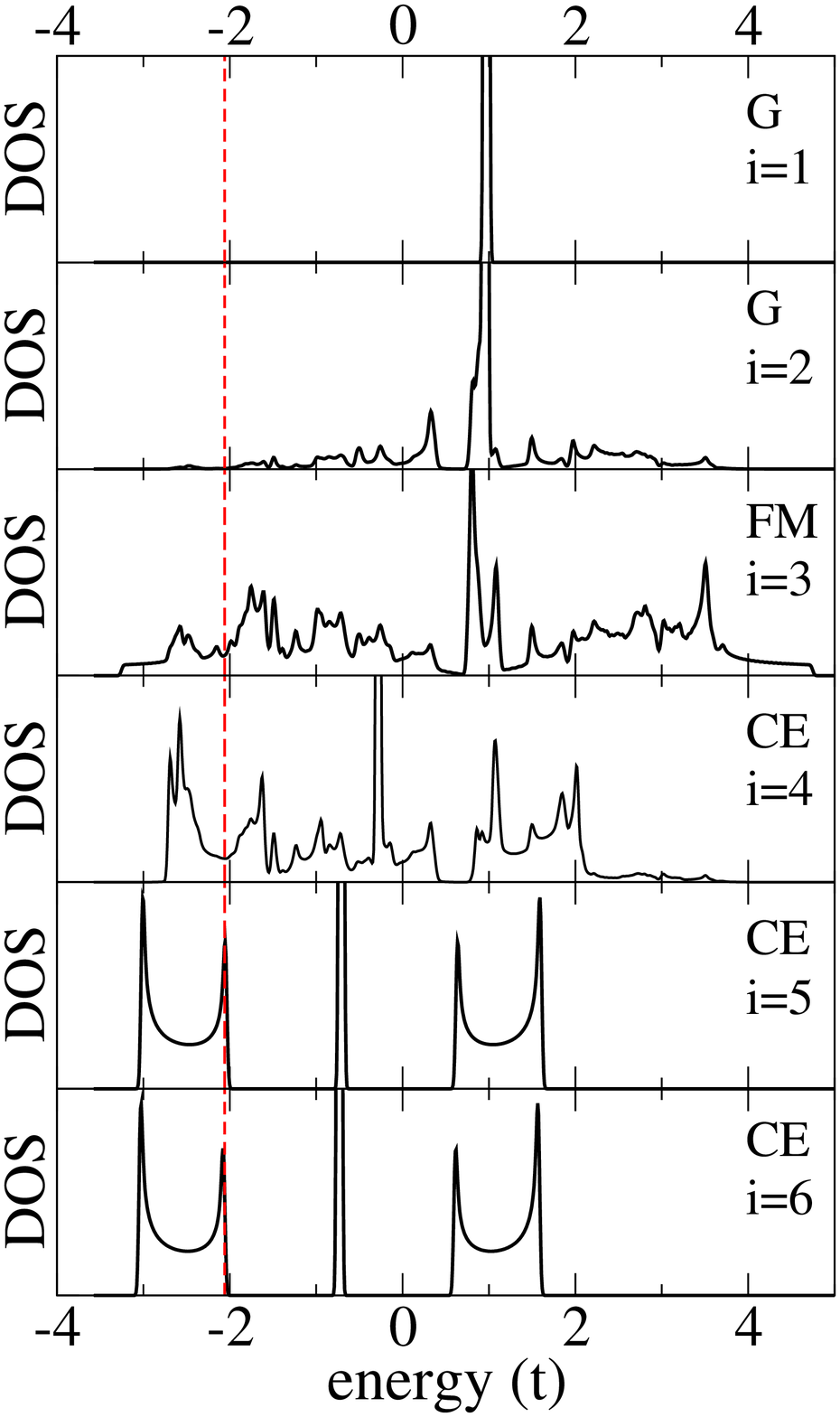}}
\caption{\label{fig:dos} (Color online) Density of states (DOS) on each plane of the heterostructure (planes $1-6$) in the $1$FM configuration for $n=0.5$ and $\alpha=2$. The vertical lines signal the position of the Fermi energy. In the center of the LCMO layer ($i=5,\, 6$) we recover the bulk CE phase [see Fig~\ref{fig:dosperfect}(a)] which is an insulator, while the CE planes closer to the interface have states at the Fermi energy. The DOS is discrete, calculated with a large number of k-points, and the curves are smoothed out by giving a gaussian weight to the eigenvalues.}
\end{figure}

We analyze now the properties of the ground state $1$FM configuration by looking at the density of states (DOS), the in-plane conductance, and the charge distribution on each of the planes. As a reference, let us first discuss the density of states of a 2-dimensional CE system illustrated in Fig.~\ref{fig:dosperfect} (a). For $n=0.5$ the first band is completely full and the system is an insulator.  For $n>0.5$, the extra electrons would have to go to the non-dispersive band at energy$=0$, but this requires a large energy, of the order of the gap, so the system would probably tend to phase separate with FM regions  (this is the case of Pr$_n$Ca$_{1-n}$MnO$_3$~[\onlinecite{jirak85}]). This phase separated system would also be insulating. For $n<0.5$, assuming the CE order is not altered, the first band would be partially full, there would be a finite DOS at the Fermi energy, and the system would be conducting. However, the experimental observation is somewhat different:  for $n<0.5$ it is possible that an incommensurate charge-orbital order arises opening a gap at the Fermi energy,~\cite{calderon-nat05,brey-PRL05} leading to insulating behavior. The DOS of a 2-dimensional FM with two e$_g$ orbitals has the peculiar shape shown in Fig.~\ref{fig:dosperfect} (b).

\begin{figure}
\resizebox{70mm}{!}{\includegraphics{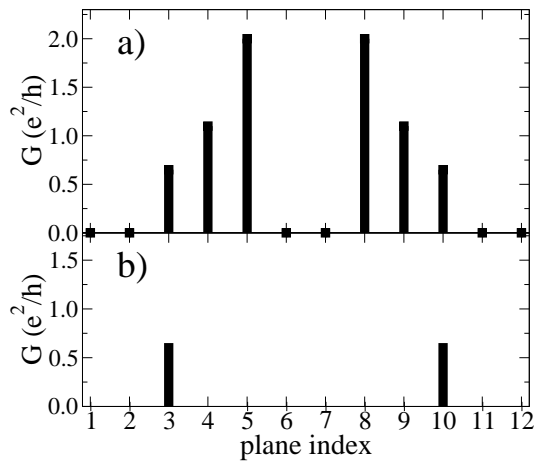}}
\caption{\label{fig:G} (a) Conductance versus the plane index for $n=0.5$. The finite conductance on the CE layers ($i=4,5,8$, and $9$) is caused by our approximation of fixing the CE ordering on those planes. (b) In fact, we only expect the FM plane to conduct (see Sec.~\ref{subsec:analysis} for discussion).
}
\end{figure}

In Fig.~\ref{fig:dos} we show the DOS on planes $i=1$ to $6$ of the CMO/LCMO heterostructure in the $1$FM configuration. The Fermi energy (E$_F$) is signaled by the dashed vertical line crossing the six panels. The shape of the DOS of a perfect CE (as in Fig.~\ref{fig:dosperfect} (a)) is recovered in planes $i=5$ and $6$, except for an overall shift due to the Hartree potential [Fig.~\ref{fig:cpl-0FM-1FM} (b)]. The electronic density in plane $i=6$ is exactly $0.5$ (see Fig.~\ref{fig:cpl-0FM-1FM}), the E$_F$ is at the gap,  and the plane is an insulator. On the other hand, the electronic density in plane $i=5$ is slightly smaller than $0.5$, there is a finite DOS at E$_F$ and, therefore, this CE plane conducts. The CE plane adjacent to the FM plane ($i=4$) has a DOS very different from that of a perfect CE, in particular, the gap at the E$_F$ closes and the plane conducts. The FM plane ($i=3$) shows no gaps, as expected.
We have also performed numerical calculations via the Kubo formula~\cite{verges,cond} of the in-plane conductance that quantify these observations. In particular, by connecting the planes one by one to the leads, we have calculated each plane's conductance.  The results are shown in Fig.~\ref{fig:G} (a) and show that three planes ($i=3$, $4$, and $5$) conduct. However, following the discussion in the previous paragraph, it is clear that the finite conductance we find in the CE planes is an artifice of our approximation of having fixed the CE magnetic ordering on the LCMO planes: it is possible (and consistent with experimental observations) that other complicate magnetic and charge-orbital orders or phase separation arose, leading to an insulating behavior. Exploring these other possibilities is out of our computing capabilities and we expect that imposing an insulating character on the doped CE phases would not affect the electrostatics of the system. The most reasonable interpretation of our results is illustrated in Fig.~\ref{fig:G} (b) where only the finite conductance in the FM plane is kept. 

\begin{figure}
\resizebox{85mm}{!}{\includegraphics{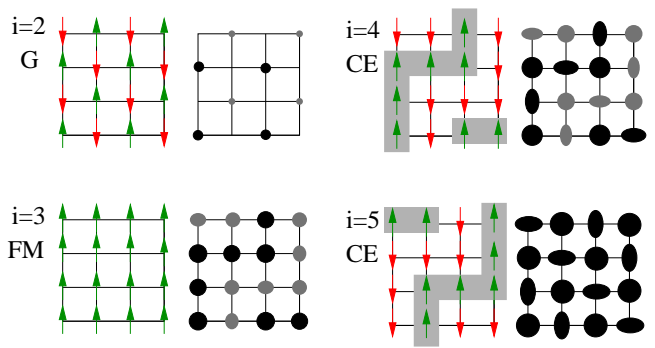}}
\caption{\label{fig:shadow} (Color online) Qualitative illustration of the magnetic ordering and charge distribution in different xy planes of the heterostructure for $n=0.5$ in the $1$FM configuration. Black and gray dots represent points with more and less charge respectively. Elongated dots represent orbital order. The middle CE planes ($i=6$-$7$, not shown) behave as in bulk: they have homogeneous charge ($0.5$ per site) and orbital ordering that distinguishes the bridge from the corner sites in the zig-zag chains.  They are also orbital polarized $x^2-y^2$ because hopping is completely suppressed in the z-direction. Plane $5$ also has homogeneous charge although $<0.5$. Plane $4$ develops charge modulation with more charge on the zig-zag chains parallel to the FM plane ($i=3$). Plane $3$ does not have charge homogeneity, although it is FM, due to the gain in kinetic energy with the sites on the parallel zig-zag chains in plane $4$. There is also a trace of orbital ordering in this FM plane, which mirrors the orbital ordering on the adjacent CE plane.
Plane $2$ gets some charge at the sites with spin parallel to the neighboring FM plane (only on the $3z^2-r^2$ orbitals). 
}
\end{figure}

Another important observation is that the DOS of the FM plane at the interface ($i=3$) is very different from that of an isolated 2-dimensional ferromagnet [see Fig.~\ref{fig:dosperfect} (b)]. This is a manifestation of the influence produced by the adjacent G and CE planes. This strong influence is also revealed by the appearance of a charge modulation ('shadow' order) on the FM plane in the heterostructure, illustrated in Fig.~\ref{fig:shadow}. Here we show both the magnetic ordering (G, FM, and CE) and a  qualitative representation of the charge and orbital configurations for planes $i=2$, $3$, $4$, and $5$. Black dots correspond to sites with more charge than in gray dots. The elongated dots represent orbital order. The charge modulation, which is absent in plane $i=5$, arises in planes $3$ and $4$ due to the fact that some sites have neighbors with a parallel spin in the z-direction while others have antiparallel neighbors. Double exchange mechanism implies that hopping is not allowed in the case of antiparallel spin and, consequently, the charge is larger at those sites with more parallel neighbors. 
 
\begin{figure}
\resizebox{75mm}{!}{\includegraphics{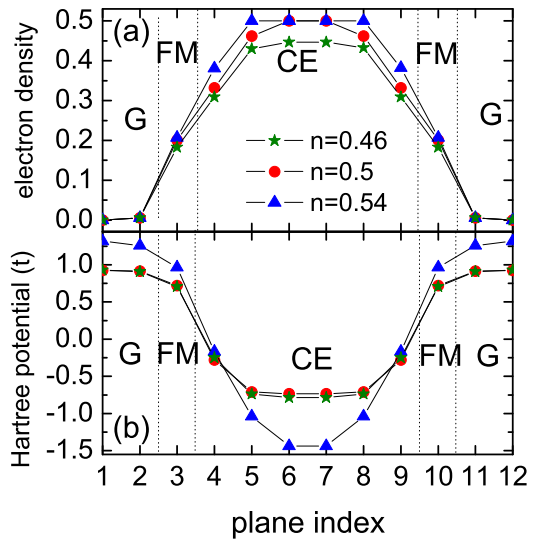}}
\caption{\label{fig:charge-potential} (Color online)  (a) Electron density per plane in the $1$FM configuration for $\alpha=1.75$ and three different values of $n$. (b) Corresponding Hartree potential. Due to the form of the density of states of a perfect CE plane [see Fig.\ref{fig:dosperfect}(a)] it is very expensive to add electrons to $n=0.5$ so the charge goes to the planes with an electron density $<0.5$. 
}
\end{figure}

\subsection{Selective doping on the LCMO layer}   

We now explore the effect of doping the LCMO layer to produce a small increase or decrease of electronic density $\delta$  (electron or hole doping, respectively). The considered heterostructure is still the same as in Fig.~\ref{fig:lattice} but now the charge on the center slab is $n=0.5 \pm \delta$ [$(0.5\pm \delta)/2$ at the interfaces]. The phase diagrams for the cases $\delta \neq 0$ are very similar to the one for $n=0.5$ shown in Fig.~\ref{fig:pd}. However, due to the asymmetry~\cite{calderon-nat05,brey-PRL05} on the electron or hole doping around $n=0.5$, manifest on the shape of the DOS in Fig.~\ref{fig:dosperfect} (a), we expect, and indeed observe, a qualitatively different behavior for $n >0.5$ and $n <0.5$ in terms of the charge distribution. 

In Fig.~\ref{fig:charge-potential} we show the electron density and Hartree potential for $n=0.46$, $n=0.5$ and $n=0.54$. For $n=0.54$, the electrons on excess of $0.5$ cannot be accommodated in the center of the LCMO layer due to the gap in the DOS [Fig.~\ref{fig:dosperfect}] so they go to the LCMO planes close to the interface and to the FM plane. 
On the contrary, the defect of electrons for the case $n=0.46$ easily accommodates at the center of the LCMO layer and there is an overall decrease of the electron density at all planes. Consequently, the electron density in the FM plane increases with increasing $n$.

\begin{figure}
\resizebox{75mm}{!}{\includegraphics{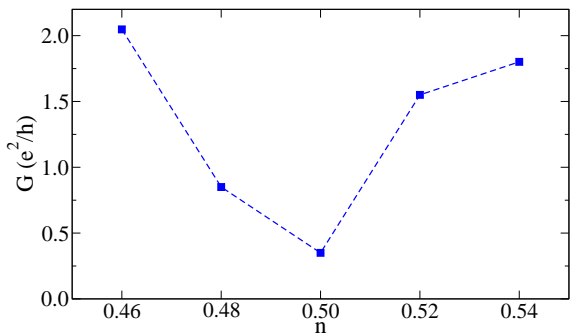}}
\caption{\label{fig:G-vs-n} (Color online) In-plane conductance of the FM plane versus $n$ around half-doping for $\alpha=2$. The overall shape of this curve depends on the value of $\alpha$.
}
\end{figure}

The in-plane conductance $G$ in the FM plane as a function of $n$ around half-doping is shown in Fig.~\ref{fig:G-vs-n} for $\alpha=2$. For this particular value of $\alpha$, the conductance increases with both electron and hole doping around $n=0.5$. However, the shape of the curve is different for different values of $\alpha$. This variability is due to the complicated structure (with many maxima and minima) of the DOS in the FM plane (see Fig.~\ref{fig:dos}). The structure on the DOS arises from the interplay between the DOS of a perfect CE [Fig.~\ref{fig:dosperfect} (a)], that of 2-dimensional FM [Fig.~\ref{fig:dosperfect} (b)], and that of a G plane (single peak at energy$=0$), shifted relatively to each other by the Hartree potential. Changing $n$ and/or $\alpha$ produce a change in both the Fermi energy and the Hartree potential so the peaks on the DOS move with respect to each other. As the conductance $G$ is a measure of the DOS at E$_F$, it is not surprising that small changes in $n$ and $\alpha$ can produce relatively large changes in $G$.  
 
\section{Conclusions}
\label{sec:conclusion}
The possibility of growing good quality oxide heterostructures is opening a new playground to study both fundamental aspects of strongly correlated systems and the competition between their different phases, and potential applications of the new phases that arise at the interfaces. Manganites, with their very rich phase diagrams, are excellent candidates to analyze all these issues.   

Here we have studied an all-manganite heterostructure consisting of a band insulator (CaMnO$_3$) and a half-filled ordered manganite (La$_{0.5}$Ca$_{0.5}$MnO$_3$) in which the insulating behavior is driven by strong correlations in the system. We use a model which has been proved successful in reproducing the phase diagram of bulk manganites.~\cite{vandenbrink-PRL99} The Coulomb interaction between the charges in the system, introduced at a Hartree level, produces a spreading of the charge across the interface. This leads to the formation of a metallic and ferromagnetic plane at the interface, similarly to what has been observed in heterostructures of Mott and band insulators,~\cite{okamoto-nat04} for a wide range of physical parameters. We have characterized this interface by its density of states and the in-plane conductance. The ferromagnetic metal at the interface shows charge and orbital modulation due to the interaction with the ordered neighboring planes. Finally, we have also observed that small electron and hole doping in the La$_{0.5}$Ca$_{0.5}$MnO$_3$ layer can produce large changes in the conductance of the ferromagnetic metallic interface.  

This work is supported by MAT2006-03741 (MEC, Spain). J.S. also acknowledges FPU program (MEC, Spain) and M.J.C. Ram\'on y Cajal program (MEC, Spain).

\bibliography{manganites}

\end{document}